\documentclass[iop]{emulateapj}
\usepackage{graphicx}
\usepackage{graphics}
\usepackage{amsmath}
\usepackage{multirow}
\usepackage{color}
\usepackage{xcolor}

\usepackage{float}

\def\kms{{\rm\,km\,s^{-1}}}
\def\kmskpc{{\rm\,km\, \,s^{-1} \, {kpc}^{-1}}}

\def\Gyr{{\rm\,Gyr}}

\def\kpc{{\rm\,kpc}}
\def\pc{{\rm\,pc}}

\def\mathnew{\mathsurround=0pt}   
\def\simov#1#2{\lower .5pt\vbox{\baselineskip0pt  
    \lineskip-.5pt\ialign{$\mathnew#1\hfil##\hfil$\crcr#2\crcr\sim\crcr}}}

\def\'#1{\ifx#1i{\accent"13\i}\else{\accent"13#1}\fi}

\begin{document}

\shorttitle{On the Origin of High-Altitude Open Clusters in the Milky Way} \shortauthors{Martinez-Medina et al. 2016}

\title{On the Origin of High-Altitude Open Clusters in the Milky Way}

\author{L.A. Martinez-Medina $^{1}$, B. Pichardo$^{1}$,
  E. Moreno$^{1}$, A. Peimbert$^{1}$ \& H. Velazquez$^{2}$}
\affil{$^{1}$Instituto de Astronom\'ia, Universidad Nacional Aut\'onoma de M\'exico, A.P. 70--264, 04510, M\'exico, D.F., M\'exico; \texttt{lamartinez@astro.unam.mx}}

\affil{$^{2}$Instituto de Astronom\'ia, Universidad Nacional Aut\'onoma de M\'exico, Apartado Postal 877, 22860 Ensenada, B.C., M\'exico}

\begin{abstract}
We present a dynamical study of the effect of the bar and spiral arms
on simulated orbits of open clusters in the Galaxy. Specifically, this
work is devoted to the puzzling presence of high altitude open
clusters in the Galaxy. For this purpose we employ a very detailed
observationally motivated potential model for the Milky Way and a
careful set of initial conditions representing the newly born open
clusters in the thin disk. We find that the spiral arms are able to
raise an important percentage of open
clusters (about one sixth of the total employed in our simulations,
depending on the structural parameters of the arms) above the Galactic plane to heights beyond
200 pc, producing a bulge-shaped structure toward the center of the
Galaxy. Contrary to what was expected, the spiral arms produce a much
greater vertical effect on the clusters than the bar, both in quantity
and height; this is due to the sharper concentration of the mass on
the spiral arms, when compared to the bar. When a bar and spiral arms
are included, spiral arms are still capable of raising an important
percentage of the simulated open clusters through chaotic diffusion
(as tested from classification analysis of the resultant high-z
orbits), but the bar seems to restrain them, diminishing the elevation
above the plane by a factor of about two.

\end{abstract}


\section{Introduction} \label{sec:intro}


The total number of open clusters in the Galaxy is estimated to be up
to one hundred thousand
\citep{2010ARA&A..48..431P,2006A&A...445..545P}; due to large amounts
of reddening, crowding, and disruption, we only know a few thousand of
them toward the Galactic center
\citep{2002ASPC..273....7V,1970ApJ...160..811F}. The ones we observe
are mostly located between 5 and 20 kpc from the Galactic center, with
ages ranging from a few Myr to approximately 10 Gyr. From the total,
only about 10\% survive their embedded stage as gravitationally bound
systems \citep{2003ARA&A..41...57L}. Orbital elements of open clusters
show that the majority were formed within a galactocentric radius of
10.5 kpc and closer than 180 pc from the Galactic plane
\citep{2012AstL...38..519G,2002A&A...389..871D}. The most massive open
clusters can reach lifetimes comparable to the age of the universe,
probably even becoming the seeds of some globular clusters
\citep{2010ARA&A..48..431P,1995ARA&A..33..381F}.

Open clusters are thought to form from dense molecular gas clouds.  In
the Galaxy molecular gas is at an extremely short scale-heigth of the
disk plane, approximately 50 to 75 pc
\citep{1994ApJ...433..687M,1998A&A...340..543V,1999A&A...344..955W},
and moves in dynamically cold orbits (almost circular and with low
inclination), therefore it is naturally expected open clusters to
follow these type of orbits. Although some molecular gas is known to
exist above the Galactic plane, these high-latitude clouds, are orders
of magnitude smaller than giant molecular clouds and they are
frequently not self-gravitating \citep{1999ASIC..540....3B}. However,
there is a non-negligible fraction (13\%) of open clusters at heights
greater than 200pc over the plane of the Galaxy, far beyond where the
molecular clouds reside \citep{2008ApJ...685L.125D}, which might be an
indication either of an unusual origin
\citep{2010MNRAS.407.2109V,2009MNRAS.399.2146W}, or of some dynamical
effect produced by the Galaxy, as discussed in this letter.

Open clusters older than 1 Gyr seem to be part of a separate structure
gravitationally trapped by the main body of the Galaxy known as the
``cluster thick disk''
\citep{2006AJ....131.1559V,2012AstL...38..519G}.

Explaining how open clusters reach such altitudes over the Galactic
plane poses a difficult challenge since no universally accepted
explanation has been found for their existence. The preferred
scenarios that have been proposed to address this problem postulate that clusters are
captured from satellite galaxies or formed through star formation
events in situ of high Galactic altitude gas clouds
\citep{2013MNRAS.434..194D,2008ApJ...685L.125D,1999ApJ...526L..89M,1977MNRAS.180..709W}. 
However, the Galaxy seems to be dominated by smooth and well-mixed
vertical age and metallicity gradients, meaning that the main
mechanisms governing its evolution have been for a long time internal
dynamical processes \citep{2015arXiv151001376C}, this provides us with
a hint to select for the present work, a deeper study on the
non-axisymmetric large-scale structures as the drivers for the
presence of a fraction of the high-altitude open clusters. This
mechanism has been already proposed and disregarded in the past on the
basis of theoretical simplified considerations such as direct (and of
low probability) close encounters of clusters with compact molecular
clouds employing the impulse approximation, or with the spiral arms as
a potential well using idealized models
\citep{1958ApJ...127...17S,Wielen1977}. However, with
three-dimensional extensions, \citet{Quillen2002} and
\citet{Quillen2014} found that vertical resonance capture might be a
mechanism to lift particles above the plane.

\section{Methodology and Numerical Implementation}
\label{model}
To carry out this study, a careful construction of the initial
conditions was a key ingredient, in addition to a very detailed,
observationally motivated Milky Way galactic potential. The open
clusters are represented as test particles in the multicomponent
galactic field. In this approximation we are interested only on the
destiny of clusters, each as a unit, from an orbital point of view.

\subsection{Milky Way's Gravitational Potential}
The model includes an axisymmetric potential, formed by a Miyamoto \&
Nagai (1975) disk and bulge, and a massive halo \citep{Allen1991}.
For the spiral arms we employ the PERLAS model \citep{PMME03} that
consists of a bisymmetric self-gravitating three-dimensional density
distribution. For the Galactic bar we use a non-homogeneous triaxial
ellipsoid that reproduces the density law of the COBE/DIRBE triaxial
central structure of the Galaxy. For further details on the model see
\citet{PMME03,Pichardo2004}.

To fit our model we make use of observational/theoretical parameters
in literature. The triaxial length of the bar is set to 3.5 kpc, with
scale-lengths of 1.7, 0.64 and 0.44 kpc. The total mass is 1.4 $\times$
10$^{10}$ M$_{\odot}$ with a pattern speed $\Omega_B$ = 45
$\kmskpc$. For the spiral arms, we consider a pitch angle of $i$ =
15.5$^{\circ}$ and a mass ratio of $M_{\rm arms}/M_{\rm disc}$ = 0.05
at a pattern speed $\Omega_S$ = 20 $\kmskpc$. Further details on the
parameters of the model and restrictions were introduced in
\citet{PMA12}.

\subsection{Cluster's Initial Conditions}
\label{sec:IC}
The initial conditions for the particle distribution are setup by
following the Miyamoto-Nagai density profile for the disk.


To discretize the Miyamoto-Nagai density profile and distribute
particles, we used the Von Neumann accept/reject algorithm
\citep{Press92}. Particle velocities are assigned with the strategy
proposed by \citet{Hernquist1993}, approximated by the second moment
of the Boltzmann equation assuming the epicyclic approximation.

We ran control simulations with the axisymmetric components of the
model to check that the initial velocity dispersion of the particle
distribution does not evolve significantly over 5 Gyrs, this ensures
that the initial stellar disk particle distribution is relaxed within
the background potential.

\subsection{Introducing adiabatically the spiral arms and the bar}
\label{sec:SimulationsA}
We now introduce the spiral arms and the bar to the potential. Both
components are grown adiabatically into the simulation, during a period
of time long enough to avoid artificial effects in the particle
kinematics. The mass of each component will be zero at time $t = 0$
and will increase with time until they reach its maximum value at a
given time $t_i$; once $t = t_i$ the mass of each component remains
fixed. The functional form to model the mass increment with time is
taken from \citep{2000AJ....119..800D}. This particular temporal
dependence allows a smooth variation for the value of the mass within
the time interval $0 < t < t_i$. The mass of the spiral arms is given
as a fraction of the mass of the disk and the mass of the bar is taken
from the mass of the bulge. 


Following this procedure we made some tests using different growing
periods for the non axisymmetric structures $t_i = 0.5, 1, 2$ Gyr. For
each case all measurements start after the non-axisymmetric components
have been fully grown. We found no significant differences in the
final results. Nonetheless we took the value $t_i = 2$ Gyr that will
guarantee the absence of artificial effects in the measurements.

\section{Results}

\subsection{The Contribution of the Spiral Arms to the Origin of High-altitude Open Clusters}
We first isolate the spiral arms effect. We start with $10^5$
particles, representing the open clusters, distributed on a cold disk,
with $\sigma_z \leq 10\kms$ for $R \geq 4\kpc$, we follow the
evolution of the system for $5\Gyr$ after the arms have grown
completely.

\begin{figure*}
\begin{center}
\hspace{-1cm} \includegraphics[width=19cm]{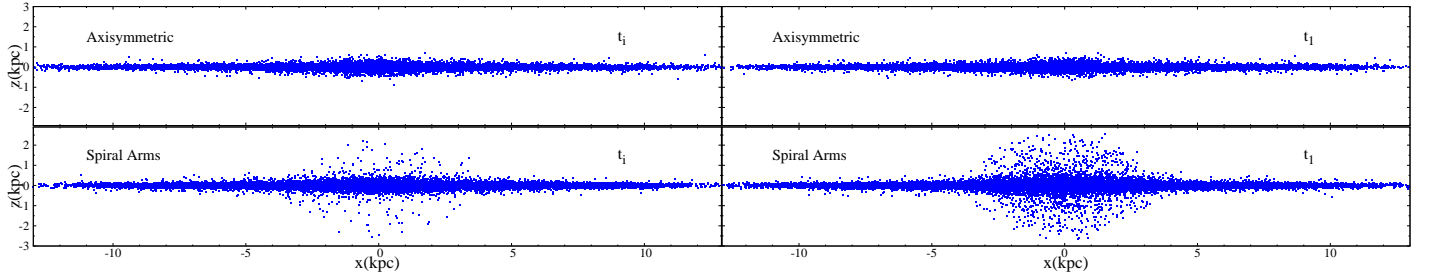}
\end{center}
\caption{$x - z$ projection of the particle disk within the
  axisymmetric background (top), and within the model with spiral arms
  (bottom), in a $1\Gyr$ lapse from $t_i$ (where spiral arms reach
  their total mass in the bottom panels) to $t_1$.}
\label{fig:control}
\end{figure*}

Figure \ref{fig:control} shows how particles in the cold disk move
away from the midplane due to the gravitational interaction with the
spiral arms. The plot is a comparison with the axisymmetric case (top
panels), along $1\Gyr$ (from left to right), where no particles are
scattered in the vertical direction. On the bottom, the galactic model
with spiral arms, from the point where arms reach their total mass at
$t_i$ (bottom left panel), to $1\Gyr$ after, at $t_1$ (bottom right
panel). In this case, lots of particles are scattered to large
distances above and below the midplane. By comparing both models, we
can establish that the collective gravitational effect of the spiral
arms induces long vertical excursions of open clusters. Note the
bulge-like structure that the model predicts would be formed by a number of
these clusters.

To quantify the number of particles departing from the initial cold
disk, first we divide the disk into radial bins of width $0.1\kpc$ and
count the number of particles with $|z| > 200\pc$.  Top left panel in
Figure \ref{fig:200pc} shows the number of particles with $|z| >
200\pc$ for each radial bin; the initial measure (blue) is taken once
the arms have reached their total mass, and from there the count is
followed for another $5\Gyr$. More and more particles raise at all
radii, although it is more efficient toward the galactic bulge.

Bottom left panel in Figure \ref{fig:200pc} shows the difference $N -
N_0$ measured every $1\Gyr$ within each radial bin. Although there are
some negative values for the quantity $N - N_0$, these are small, and the
overall trend is clear and indicates that as an outcome of the
interaction with the spiral arms, the number of particles that move
away from the galactic plane ($|z| > 200\pc$) increases with time.

\subsubsection{Alternative numerical approach}
\label{sec:SimulationsB}
With the previous procedure to introduce the spiral arms, the count of
uprising particles might be underestimated; namely, we are letting the
arms grow adiabatically for $2\Gyr$ and we are not measuring any
quantity during this period. Although starting the analysis from the
time the arms have grown completely avoids any transient effect appearing 
in the results, this procedure leave us without knowledge of the
true initial time from which the kinematics of the particles starts
being affected by the spiral arms. This means that we are losing some
information because aside from possible relaxation effects, the
particles are already interacting with the arms while these are
growing.

With the aim of recovering the physical information lost during the
growing period, we apply the following numerical procedure.

a) We distribute $10^5$ particles in density and velocity space as in
Section \ref{sec:IC}, but initially setting $z=0$ and $v_z=0$, i.e.,
we start with a two-dimensional particle distribution. Then the
simulation starts running with the arms introduced adiabatically
within the first $2\Gyr$.

b) Once the arms have reached their total mass, the numerical
integration goes for another $3\Gyr$ to allow the particles to relax
with the spiral background potential. At this point the arms are
totally formed and the stellar disk is relaxed.

c) Now for each particle at a given radial coordinate $R$, a $z$
coordinate and $v_z$ velocity are assigned following the usual
procedure (Sec. \ref{sec:IC}).  At this stage we have recovered our three-dimensional
disk but with the particles already relaxed in the radial
direction. This point is closer to what we call an ``initial time''
since at the end of this new procedure, we have gotten rid of most
of the spurious reaction, providing us with a better starting point for the
calculations. 

Finally, the integration restarts from here for another $5\Gyr$, which
includes all the periods of interest. Measurements made from here will
capture as many as possible of the particles that undergo large
vertical excursions above and below the midplane.

\begin{figure*}
\begin{center}
\includegraphics[width=18cm]{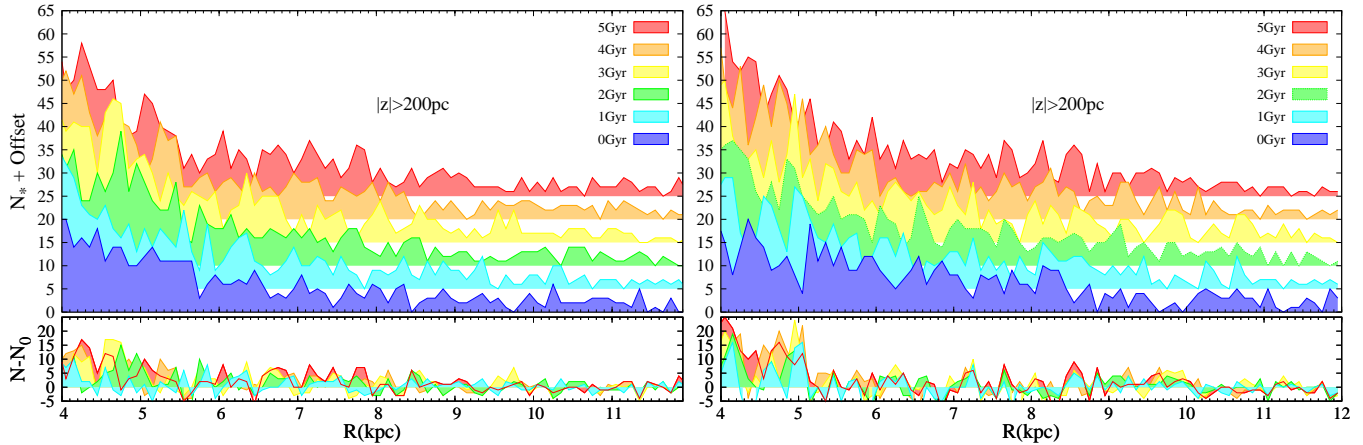}
\end{center}
\caption{(Top) Right and left panels: number of particles $N$ that at
  the indicated time have $|z|>200\pc$ for simulations described in
  Sections \ref{sec:SimulationsA} and \ref{sec:SimulationsB}
  respectively. (Bottom) Right and left panels: difference between the
  number of particles $N$ and the initial number of particles $N_0$,
  computed every $1\Gyr$ for simulations described in Sections
  \ref{sec:SimulationsA} and \ref{sec:SimulationsB} respectively.}
\label{fig:200pc}
\end{figure*}

For this alternative numerical implementation we repeated the counts
as described above for particles with $|z|>200\pc$ (right panels in
Figure \ref{fig:200pc}). We see the same behavior found before, that
the number of particles with high $|z|$ increase with time due to
their interaction with the spiral arms. But a first comparison of the top
panels in Figure \ref{fig:200pc}, shows that for the second procedure,
the number of particles at the starting point with high $|z|$, is
smaller than in the previous procedure, mainly toward the Galactic
Center. The relaxation strategy of the disk avoids the loss of
information that is inherent to the growing period of the arms. This has
direct consequences for the effective number of particles scattered
away from the midplane at the end of the simulation, as seen by
comparing bottom panels in Figure \ref{fig:200pc}. We see that the
count of uprising particles is slightly larger for the second
numerical procedure, as expected.

\subsection{The Contribution of the Bar and the Bar+Arms to the Origin of High-altitude Open Clusters}
We have shown that particles in a cold disk can be scattered away from
the midplane of the disk in an effective fashion by gravitational
interaction with the spiral arms. Now, in order to study the effect of
the galactic bar, we introduce a triaxial potential, observationally
motivated (see Section \ref{model}). After the bar is totally formed,
we follow the time evolution of the particle disk for another $5\Gyr$.

\begin{figure*}
\begin{center}
\hspace{-1cm} \includegraphics[width=19cm]{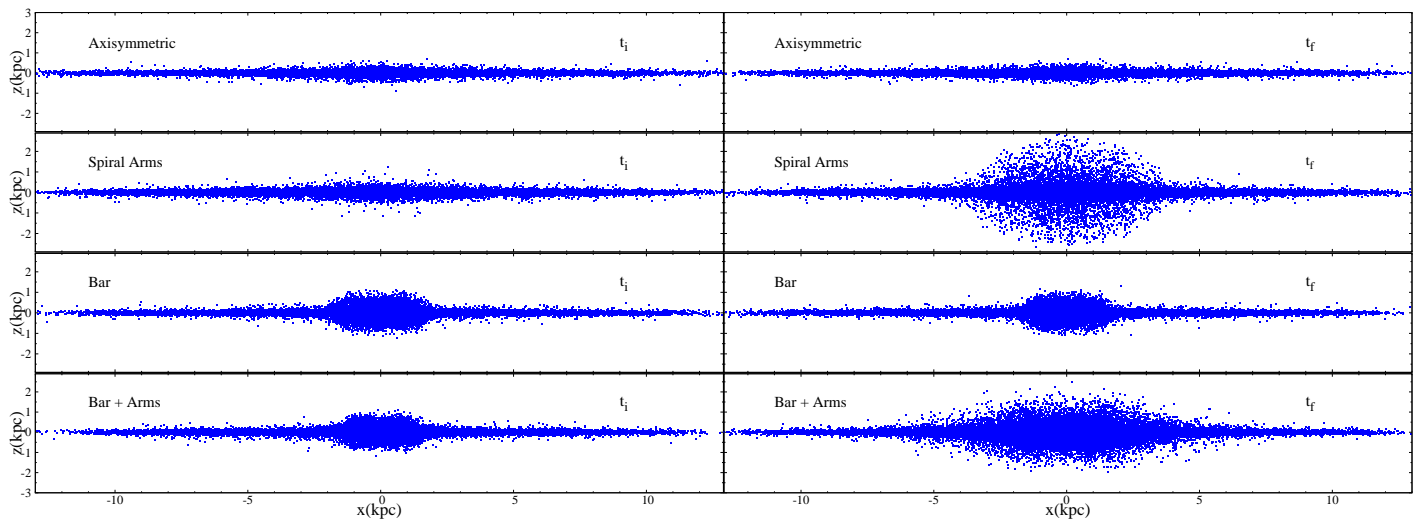}
\end{center}
\caption{$x - z$ projection of the particle disk at $t_i$ (left
  panels) and $t_f = 5\Gyr$ (right panels) within our four
  experiments: axisymmetric (first row), spiral arms (second row), bar
  (third row), and bar+arms (fourth row).}
\label{fig:2edgeon}
\end{figure*}

In the third row of Figure \ref{fig:2edgeon}, we show the initial and
final stages for the $5\Gyr$ evolution. Contrary to the result in the
previous experiment with only spiral arms, the bar does not drive
particles away from the midplane of the disk. This is because the bar,
compared to the arms, is efficient for capturing particles in orbits mostly
confined within the triaxial structure, rather than crossing
the potential well of the bar periodically, as it seems to be occurring
with the spiral arms. Therefore, the massive bar acts more as an
attractor that confines the clusters, spreading them radially instead
of scattering them away from the plane of the disk. The spiral arms, on
the other hand, are capable to imprint vertical accelerations to the
particles due to the concentration, but they do not seem to be massive
enough to retain them, resulting in particles moved away from the
plane.

The most realistic case for the Milky Way is when the bar and spiral
arms are present. To study this case, both are introduced
adiabatically during the simulation (as described in Section
\ref{sec:SimulationsA}). Figure \ref{fig:2edgeon} (fourth row) shows
the result of the $5\Gyr$ evolution. Lots of particles in this case
are again scattered away from the plane, even in the presence of the
bar. The vertical response of the disk is a combination of the
previous two experiments: the arms scatter particles to large vertical
distances, although these new distances are slightly diminished by the
presence of the bar toward the Galactic Center.

\begin{figure*}
\begin{center}
\includegraphics[width=18cm]{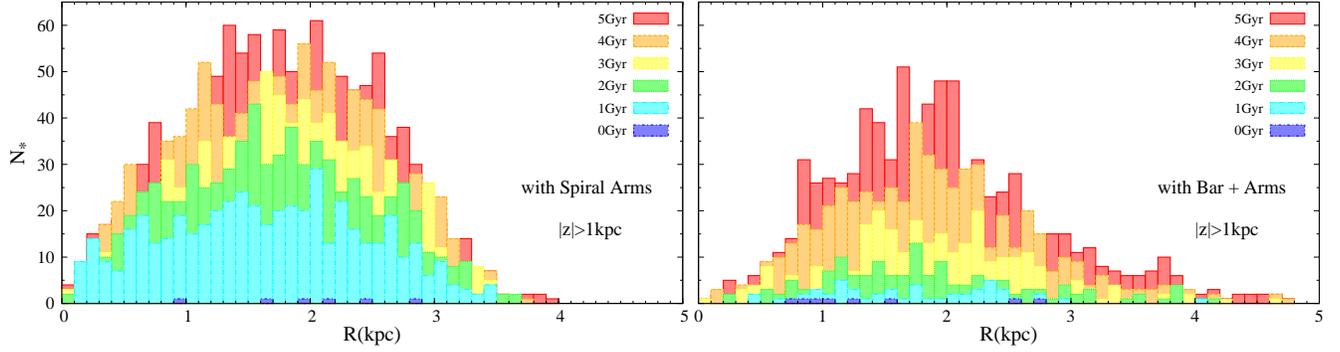}
\end{center}
\caption{Number of particles, $N$, that at the indicated time have
  $|z|>1\kpc$ in the simulation with spiral arms (left) and in the
  simulation with bar+arms (right).}
\label{fig:1Kpc}
\end{figure*}

Quantifying the effect with particle counts, we noticed they can reach
altitudes larger than $1\kpc$. By dividing the disk into radial bins
of width $0.1\kpc$ we count the number $N$ of particles with $|z| >
1\kpc$ every $1\Gyr$ as shown in Figure \ref{fig:1Kpc} for the
simulations with spiral arms (left) and bar+arms (right).  At the
initial time the number of particles with $|z|>1\Gyr$ is almost zero
for the two experiments, and from there the number $N$ of these
particles increases with time in both cases.  As we saw earlier, when
comparing the third and fourth rows in Figure \ref{fig:2edgeon},
Figure \ref{fig:1Kpc} shows that the number $N$ of particles with $|z|
> 1\kpc$ is always greater for the case with the spiral arms alone
than with the combination of bar+arms. In either case, particles are
still being scattered to high altitudes, $1\kpc$ or more above the
plane.

\subsection{Orbital analysis}
\begin{figure*}
\begin{center}
\includegraphics[width=18cm]{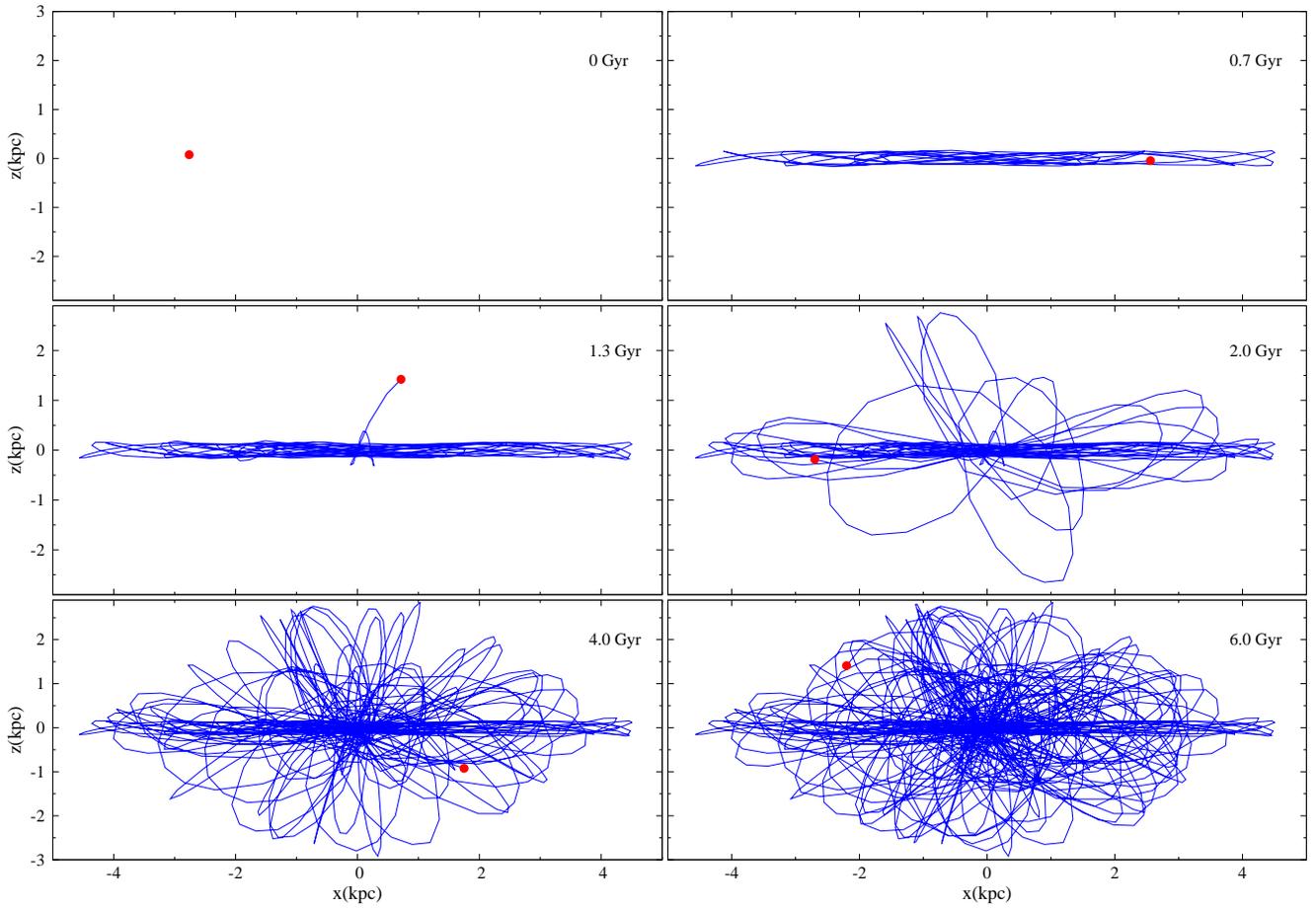}
\end{center}
\caption{Time evolution for one of the clusters which, in the
  simulation with spiral arms, ends up with $|z| > 1\kpc$. The cluster
  ascends at approximately 1 Gyr at a Galactic radius of $\sim 5$ kpc.}
\label{fig:orb}
\end{figure*}

The collective effect of the spiral arms produces a chaotic diffusion
of a number of the very cold open cluster orbits that drives them very
high over the disk plane. In Figure \ref{fig:orb} we show one example
of those orbits. At early times the particle moves mainly on the
plane, then at $1.3\Gyr$, the orbit is deflected upwards. After this
point the orbit is no longer confined to the plane; further
interactions with the spiral arms move it to even higher altitudes.

Chaotic stellar orbits are very common, and in some cases, they may even
outnumber the regular orbits in a stellar system. To see if this is
the case for the particles with high altitudes in our simulations we
used the classification code by \citet{Carpintero1998}, based on the
method of spectral dynamics introduced by
\citet{Binney1982,Binney1984}. We applied the orbital classifier to
the total sample of orbits that ended up the simulation with high
$z$. In the specific case of high-reaching orbits, we found that: one-third 
of the orbits that go above 200 pc are chaotic orbits, with the
rest being tube orbits; for the orbits that go above 500 pc, this
fraction has gone up to about one-half; by the time the orbits go
beyond 1 kpc, slightly over one half are chaotic. Chaos seems to be
the dominant mechanism to explain the highest reaching orbits. As for
the tube high-reaching orbits, the mechanism might be associated with
resonances.


\section{Conclusions} \label{conclusions}
With the use of a detailed observationally motivated model of the
Milky Way Galaxy and a set of carefully constructed cold initial
conditions for the newly born open clusters, we study the effect of
the spiral arms and bar on the orbits of open clusters. Particularly,
we are interested in the contribution of the large-scale
non-axisymmetric structures to the high elevation of open clusters
over the galactic plane.

We find that spiral arms are able to induce large excursions of
clusters' orbits up to 3 kpc heights over the plane of the disk,
depending on their specific parameters. All results in this work were
computed assuming a spiral arms mass of $5\%$ of the disk
(corresponding to a relative torque $Q_{T(max)} \sim 0.1$) and a pitch
angle of 15.5$^{\rm o}$. We performed a second simulation with a mass
of $3\%$ ($Q_{T(max)}\sim 0.06$). For the second case, the behavior is
similar but diminished by a factor of $\sim 0.7$ by number, and a
factor of $\sim 0.75$ by height, with respect to the $5\%$ mass case
of the spiral arms (in a future paper we will present this
study). This means that for spiral arms with larger pitch angles than
the 15.5$^{\rm o}$ usually assumed for the Milky Way, or higher masses
of the spiral arms for instance, the quantity of these clusters could
be even larger.

On the other hand, the bar, despite its total mass, has a much smaller
effect. The difference in this behavior is due to the density, which
allows the spiral arms to give stronger kicks to the clusters. The net
effect of the bar is to concentrate the orbits toward the galactic
plane and to produce a radially outward diffusion of cluster
orbits.

In the full model that includes both spiral arms and bar, despite the
presence of the more massive bar, spiral arms are readily able to
raise up an important percentage of the simulated open clusters
through chaotic diffusion, as tested from a spectral classification
analysis of the resultant high-z orbits; while the bar produces the
same radial diffusion and at the same time seems to concentrate the
orbits toward the Galactic plane, slightly reducing the effect of the
arms.

The cluster system forms a bulge-like structure. Although the bar shrinks 
the cluster system, it is still half (or more) of the size of the case 
with only spiral arms. In an extended ongoing paper (L.A. Martinez-Medina
et al. in preparation), we will present a detailed study based both on
an analytical approximation to destruction rates and tidal radii of
known clusters and also a second order analytical approximation and
finally N-body simulations with clusters to ponder their survival
rates in the plane and off the plane in a Milky Way detailed model.

\acknowledgments We would like to thank the referee for insightful
comments that helped to improve this work. We also acknowledge the
support of DGTIC-UNAM and Instituto de Astronom\'ia, UNAM for
providing HPC resources on the Cluster Supercomputers Miztli and
Atocatl. We acknowledge DGAPA-PAPIIT through grants IN-114114 and
IG-100115. L.A.M.M. acknowledges support from a DGAPA-UNAM postdoctoral
fellowship.


\begin{thebibliography}  

\bibitem[\protect\citeauthoryear{Allen \& Santill\'an}{1991}]{Allen1991}
Allen, C. \& Santill\'an, A. 1991, Rev. Mexicana Astron. Astrofis., 22, 256

\bibitem[Binney \& Spergel (1982)]{Binney1982} 
Binney, J., \& Spergel, D., 1982, ApJ, 252, 380

\bibitem[Binney \& Spergel (1984)]{Binney1984} 
Binney, J., \& Spergel, D., 1984, MNRAS, 206, 159

\bibitem[Blitz \& Williams(1999)]{1999ASIC..540....3B} Blitz, L., \&
  Williams, J.~P.\ 1999, NATO Advanced Science Institutes (ASI) Series
  C, 540, 3

\bibitem[Carpintero \& Aguilar(1998)]{Carpintero1998} 
Carpintero, D.~D., \& Aguilar, L.~A.\ 1998, \mnras, 298, 1

\bibitem[Casagrande et al.(2015)]{2015arXiv151001376C} Casagrande, L., 
Silva Aguirre, V., Schlesinger, K.~J., et al.\ 2015, arXiv:1510.01376 

\bibitem[de la Fuente Marcos \& de la Fuente Marcos(2008)]{2008ApJ...685L.125D} 
de la Fuente Marcos, R., \& de la Fuente Marcos, C.\ 2008, \apjl, 685,
L125

\bibitem[de la Fuente Marcos et al.(2013)]{2013MNRAS.434..194D} 
de la Fuente Marcos, R., de la Fuente Marcos, C., Moni Bidin, C.,
Carraro, G., \& Costa, E.\ 2013, \mnras, 434, 194

\bibitem[Dehnen(2000)]{2000AJ....119..800D} Dehnen, W.\ 2000, \aj, 119, 800 

\bibitem[Dias et al.(2002)]{2002A&A...389..871D} Dias, W.~S., Alessi,
  B.~S., Moitinho, A., \& L{\'e}pine, J.~R.~D.\ 2002, \aap, 389, 871


\bibitem[Freeman(1970)]{1970ApJ...160..811F} Freeman, K.~C.\ 1970, \apj, 
160, 811 

\bibitem[Friel(1995)]{1995ARA&A..33..381F} Friel, E.~D.\ 1995, \araa, 33, 381 

\bibitem[Gozha et al.(2012)]{2012AstL...38..519G} Gozha, M.~L., Koval', 
V.~V., \& Marsakov, V.~A.\ 2012, Astronomy Letters, 38, 519 

\bibitem[\protect\citeauthoryear{Hernquist}{1993}]{Hernquist1993}
Hernquist L., 1993, ApJS, 86, 389

\bibitem[Lada 
\& Lada(2003)]{2003ARA&A..41...57L} Lada, C.~J., \& Lada, E.~A.\ 2003, \araa, 41, 57 


\bibitem[Malhotra(1994)]{1994ApJ...433..687M} Malhotra, S.\ 1994, \apj, 
433, 687 

\bibitem[Martos et al.(1999)]{1999ApJ...526L..89M} Martos, M., Allen, C., 
Franco, J., \& Kurtz, S.\ 1999, \apjl, 526, L89 

\bibitem[\protect\citeauthoryear{Miyamoto \& Nagai }{1975}]{MN75}
Miyamoto, M., \& Nagai, R., 1975, Pub. Astr. Soc. Japan, 27, 533

\bibitem[\protect\citeauthoryear{Pichardo et al.}{2003}]{PMME03}
Pichardo, B., Martos, M., Moreno, E. \& Espresate, J., 2003, ApJ, 582, 230

\bibitem[Pichardo et al.(2004)]{Pichardo2004} Pichardo, B., Martos, 
M., \& Moreno, E.\ 2004, \apj, 609, 144

\bibitem[Pichardo et al.(2012)]{PMA12} 
Pichardo B., Moreno E., Allen C., Bedin L. R., Bellini A., 
Pasquini L., 2012, \aj, 143, 73 

\bibitem[Piskunov et al.(2006)]{2006A&A...445..545P} Piskunov, A.~E.,
  Kharchenko, N.~V., R{\"o}ser, S., Schilbach, E., \& Scholz,
  R.-D.\ 2006, \aap, 445, 545

\bibitem[Portegies Zwart et al.(2010)]{2010ARA&A..48..431P} Portegies
  Zwart, S.~F., McMillan, S.~L.~W., \& Gieles, M.\ 2010, \araa, 48,
  431

\bibitem[Press et al. (1992)]{Press92} Press, W. H., Teukolsky, S. A.,
  Vetterling, W. T., \& Flannery, B. P. 1992, Numerical recipes in
  FORTRAN. The art of scientific computing

\bibitem[\protect\citeauthoryear{Quillen}{2002}]{Quillen2002}
Quillen, A. C. 2002, AJ, 124 ,722

\bibitem[\protect\citeauthoryear{Quillen et al.}{2014}]{Quillen2014}
Quillen A. C., Minchev I., Sharma S., Qin Y.-J., Di Matteo P., 2014, MNRAS, 437, 1284

\bibitem[Spitzer(1958)]{1958ApJ...127...17S} Spitzer, L., Jr.\ 1958, \apj, 
127, 17 

\bibitem[Vande Putte et al.(2010)]{2010MNRAS.407.2109V} Vande Putte,
  D., Garnier, T.~P., Ferreras, I., Mignani, R.~P., \& Cropper,
  M.\ 2010, \mnras, 407, 2109

\bibitem[Van den Bergh(2006)]{2006AJ....131.1559V} van den Bergh, S.\ 2006, 
\aj, 131, 1559

\bibitem[van der Kruit(2002)]{2002ASPC..273....7V} van der Kruit, P.~C.\ 
2002, The Dynamics, Structure \& History of Galaxies: A Workshop in 
Honour of Professor Ken Freeman, 273, 7 

\bibitem[Vergely et al.(1998)]{1998A&A...340..543V} Vergely, J.-L.,
  Ferrero, R.~F., Egret, D., \& Koeppen, J.\ 1998, \aap, 340, 543


\bibitem[Wei{\ss} et al.(1999)]{1999A&A...344..955W} Wei{\ss}, A.,
  Heithausen, A., Herbstmeier, U., \& Mebold, U.\ 1999, \aap, 344, 955

\bibitem[Wielen(1977)]{Wielen1977} Wielen, R.\ 1977, \aap, 60, 263

\bibitem[Williams et al.(1977)]{1977MNRAS.180..709W} Williams, P.~M., 
Brand, P.~W.~J.~L., Longmore, A.~J., 
\& Hawarden, T.~G.\ 1977, \mnras, 180, 709 

\bibitem[Wu et al.(2009)]{2009MNRAS.399.2146W} Wu, Z.-Y., Zhou, X., Ma, J., 
\& Du, C.-H.\ 2009, \mnras, 399, 2146


\end{thebibliography}
\end{document}